# Evidence of the Monopolar-Dipolar Crossover Regime: A Multiscale Study of Ferroelastic Domains by In-Situ Microscopy Techniques

John J R Scott[1*], Guangming Lu[2*], Brian J. Rodriguez[3], Ian MacLaren[4], Ekhard K.H. Salje[5], Miryam Arredondo[1*]

[1*] School of Mathematics and Physics, Queen's University Belfast, Belfast BT7 1NN, Northern Ireland, UK

[2] School of Environmental and Material Engineering, Yantai University, Yantai 264005, China

[3] School of Physics, University College Dublin, D04 V1W8, Dublin, Ireland

[4] School of Physics and Astronomy, University of Glasgow, Glasgow G12 8QQ, UK

[5] Department of Earth Sciences, University of Cambridge, Cambridge CB2 3EQ, UK

* **jscott63@qub.ac.uk, luguangming1990@stu.xjtu.edu.cn, m.arredondo@qub.ac.uk**

**Abstract**

The elastic interaction between kinks (and antikinks) within domain walls plays a pivotal role in shaping the domain structure, and their dynamics. In bulk materials, kinks interact as elastic monopoles, dependent on the distance between walls ($d^{-1}$) and typically characterised by a rigid and straight domain configuration. In this work we investigate the evolution of the domain structure as the sample size decreases, by means of in-situ heating techniques on free-standing samples. A significant transformation is observed: domain walls exhibit pronounced curvature, accompanied by an increase in both domain wall and junction density. This transformation is attributed to the pronounced influence of kinks, inducing sample warping, where 'dipole-dipole' interactions are dominant ($d^{-2}$). Moreover, we experimentally identify a critical thickness range that delineates a cross-over between the monopolar and dipolar regimens and corroborated this by detailed atomic simulations. These findings are relevant for in-situ TEM studies and for the development of novel devices based on free-standing ferroic thin films and nanomaterials.

## Introduction

In the realm of ferroics, domains are regions with uniform atomic arrangements and order parameters such as polarisation in ferroelectric materials or magnetisation in ferromagnets. These domains are separated by interfaces known as domain walls, which are recognised as functional elements with distinct emerging properties and that can be created, manipulated, and annihilated by applying external fields. The field induced changes in the domain pattern lead to hysteretic behaviour with dynamics often characterized by avalanche dynamics.[1-3] The control over domains and domain walls, their formation and dynamics, lies hence at the heart of developing novel technologies[4] such as neuromorphic computing,[5-7] memory storage[8, 9] and telecommunications.[10, 11]

Ferroelastic domains play a crucial role in mediating many of the sought-after properties of ferroelectrics and multiferroics,[12-14] especially when strain and polarization cannot be fully uncoupled.[15] With the ongoing trend towards device miniaturisation[16-18] and the increasing interest on free-standing ferroic device elements,[19-21] a more comprehensive understanding on the domain and domain wall behaviour, and its underlying mechanisms, across different length scales has become pivotal.

Domain walls propagate with a superposition of smooth-ballistic and jerky movements[22, 23] as they repeatedly encounter defects such as oxygen vacancies. Additionally, domain walls are capable of displaying complex morphology in the form of bends and meanders, and importantly, atomic steps known as kinks.[3] Upon overcoming defects, the domain wall releases long-ranging strain fields[24] triggered by these kinks,[1, 25] which can destabilise the surrounding domain structure to form a cascade of events (avalanches).

The interaction of kinks is known to have a distance ($d$) dependency. In bulk materials, kinks interact as elastic monopoles over wall distances that are similar in distance to the average kink distribution per wall[26] with a dependency of $d^{-1}$. For thin samples, this behaviour transitions to an interacting distance between kinks of $d^{-2}$, where 'dipole-dipole' interactions are dominant. Theoretical work has elucidated a cross-over between the monopolar and dipolar regimes, suggesting a rather wide, continuous regime for samples sizes between 200-600nm based on the scaling exponents for different system sizes.[20] This cross-over regime can be rationalised in terms of the volume reduction: the warping of the surface, where each domain wall creates localised strain that leads to macroscopic bending, with the bending direction depending upon the type of kink pairing within the wall.[20, 26] Indeed, it has been demonstrated that kinks play a key role within domain switching and domain wall propagation[27] whilst demonstrating their influence on properties such as the dielectric response,[28] ferroelectricity and superconductivity which presents opportunities for novel applications[29] such as supersonic data transfer.[30] Despite these opportunities, the physical principles of these kinks and some of the more basic interactions, such as those between parallel domain walls, have only been recently reported[26] highlighting the need for further studies.

In bulk, ferroelastic domains typically present a domain pattern with a rigid and straight configuration, as their total internal energy is related to elastic forces that extend over the whole sample, unless exaggerated local strains are present. For this case, interactions between domain walls are dominated by junctions of vertically and horizontal domains which determines the functional properties of these materials.[26] Interestingly, it has been demonstrated that even within the bulk, the aspect ratio (length/width) directly influences the resulting domain structure and can induce distinct domain dynamics.[2] It can be expected that as the system size decreases, altering the aspect ratio and volume, the internal energy will reduce due to the non-local nature of elastic forces that depend on the total volume. Thus, as the dimensionality decreases, the dominant interactions (once driven by domain junctions) will greatly rely upon kinks and their interactions. In turn, the possibility of domain wall bending increases[24] and this could induce additional kinks within the walls.[30] Theoretical simulations allude[26] to changes in domain configuration under different boundary conditions i.e., the difference between clamped and freestanding samples[20] with the thinnest samples reminiscent of super-elasticity in $BaTiO_3$ membranes.[31] This results in an increased number of domain intersections and domain wall bending,[32] which have been shown to exhibit enhanced properties. However, experimental evidence on the different regimes, and their effect on the domain configuration as a function of system size, has not been fully explored.

Here, we present a systematic study on the cross-over regime by using in-situ heating optical microscopy in tandem with in-situ heating scanning transmission electron microscopy

(STEM). The resulting domain structure is analysed as a function of decreasing thickness in free-standing LaAlO$_3$, with samples ranging from 500μm to 100nm. The domain wall curvature and density were measured as the characteristic parameters between the monopolar and dipolar regimes. We identify a cross-over region based on physical observations between different domain configurations as a function of sample thickness and discuss the degree of bending of the domain walls and densities. Our experimental observations are further confirmed by 3D models adapted from previous theoretical work, clearly indicating the effect that the elastic field on the surface has on the resulting domain pattern, a result from the inherent kink-kink interactions as the thickness is reduced.

**Results**

Three samples ranging from 500μm to 220μm were investigated optically whilst heat cycled in a Linkam in-situ heating system, as described elsewhere.[2] Here, domains are on the scale of a few to tens of microns (Figure 1a-b) and scale according to Kittel's law. As expected, the domain walls are straight and rigid at this scale as predicted by Landau theory, with a hyperbolic tangent profile of the walls. Minor meandering is noted and this is rationalised as a mechanism to compensate for enhanced local strains which are further identified via the presence of herringbone domains and domains nucleated via mother-daughter kinks in agreement with previous work.[2, 30, 33, 34]

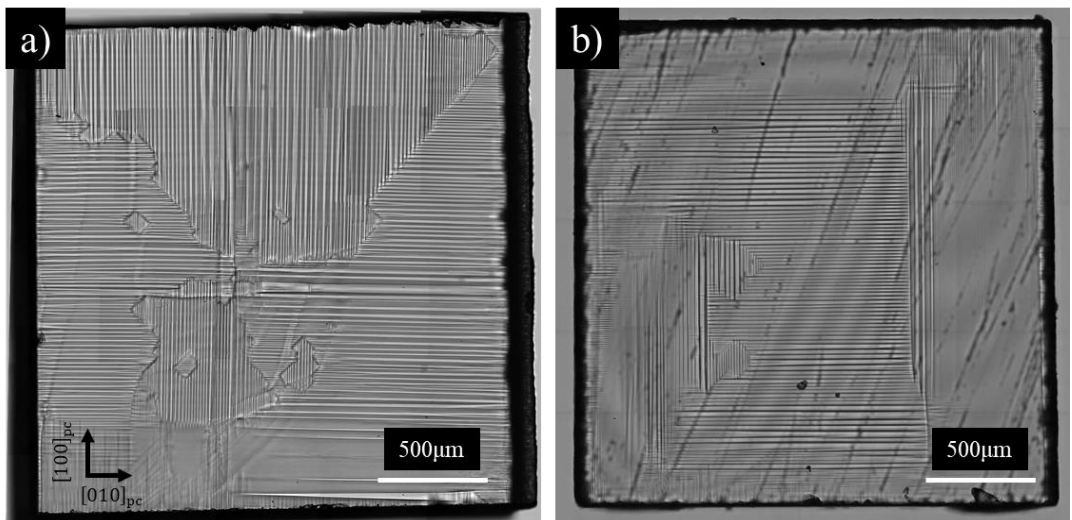

**Figure 1.** Domain structure in bulk samples. Representative optical micrographs displaying the room temperature domain structure for bulk samples, **a)** 500μm and **b)** 220μm.

To investigate the effect on decreasing the sample dimensions, 22 free-standing thin films with thicknesses ranging from 500 nm to 100 nm were fabricated as described in the methodology section. As expected, pristine samples typically demonstrated a monodomain structure with the occasional exception of one domain wall having been captured from the bulk during the fabrication process. All samples were heat cycled from room temperature to 600°C (above T$_C$ ~545°C) and back at a rate of 0.33 °C/sec, allowing the domain structure to be reset and select a more optimal configuration under the newly imposed boundary conditions. Figure 2 displays representative High-Angle Annular Dark-Field (HAADF) and Dark-Field (DF) STEM images for the samples at room temperature, after heating above $T_C$. The domain structure in the samples ranging from 500nm to 200nm exhibited domains largely parallel to the sample edges, resembling the bulk samples' domain structure. Although a change in periodicity is evident, the domain configuration was less complex and sparse (Figures 2a-b), with only a couple of junctions existing if any and the domain walls appearing relatively straight. A clear change was noted within the samples ranging between 180nm and 100nm, where an increase in the domain density and the number of junctions and/or domain wall curvature are evident.

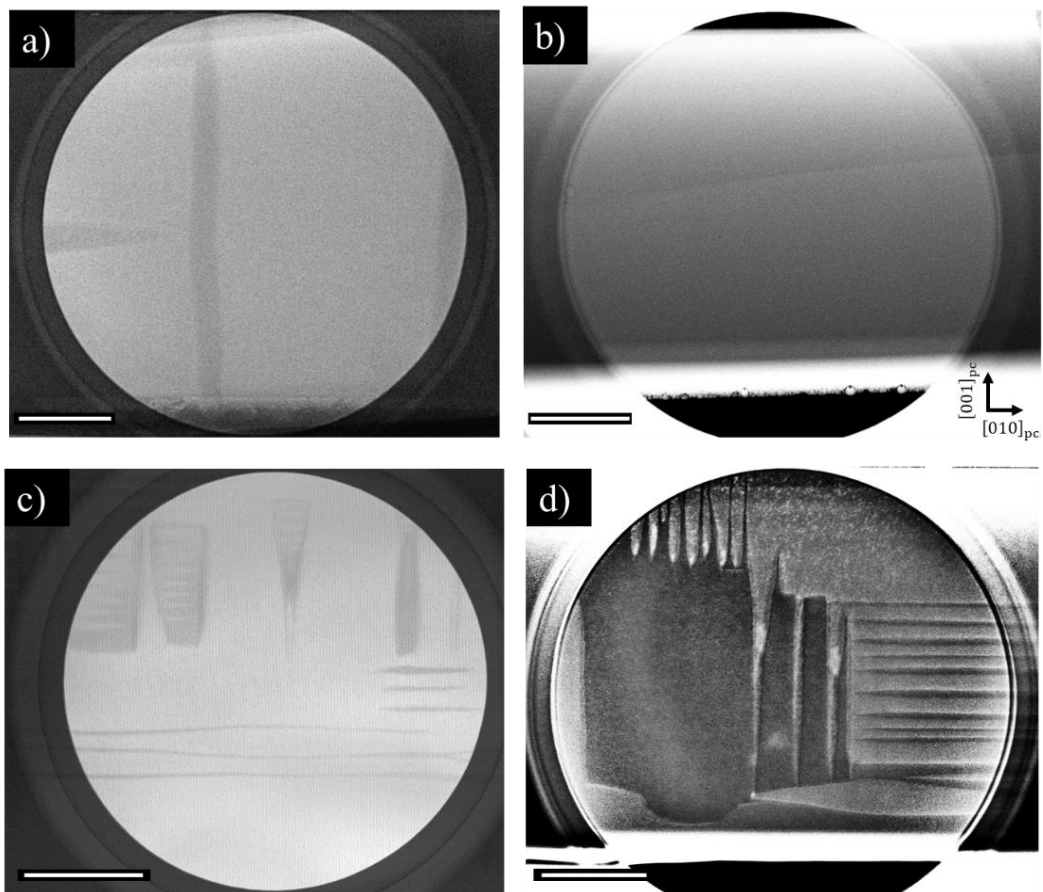

**Figure 2.** Domain structure in thin samples. Representative HAADF and DF STEM images showing the domain structure at RT for thicknesses of **a)** 400nm, **b)** 200nm, **c)** 180nm and **d)** 100nm. Between 500nm and 200nm, the domain walls are reasonably straight with a simpler domain pattern. Below 200 nm, domains walls become curvier with increased domain junction and density. The scale bar represents 2μm.

To characterise the changes, the domain wall density, junction density and domain curvature were measured as a function of sample thickness (Figure 3), for the latter the Kappa package within FIJI was used.[35] For the thickest samples (500μm to 220μm) both the domain junction and domain wall densities followed a flat trend that mirrored the trend of recorded domain curvatures (Figures 3a-b, region 3). Samples ranging from 500nm – 200nm displayed a significant increase in domain and junction density. This roughly scaled in magnitude with the rise in domain wall curvature although, the average density points begin to cluster and no longer follow a linear trend (Figures 3a-b, region 2). This deviates beyond 180nm, where the domain wall and junction densities follow a flat trend that scales with the increase to domain wall curvature (Figures 3a-b, region 1) with domain wall density following the upper limit of the curvature and junction density the lower limit.

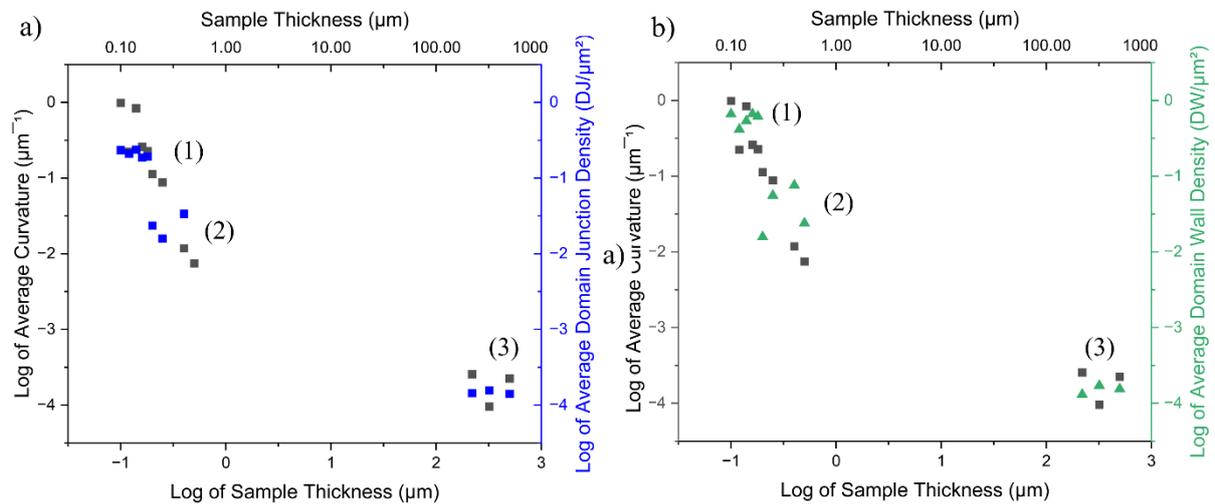

**Figure 3.** Correlation between domain wall density, junction density and curvature as a function of sample thickness. **a)** Domain junction density and **b)** domain wall density, superimposed to demonstrate their scaling to one another. From these measurements, 3 regions can be identified (marked 1-3), alluding to the existence of the monopolar-dipolar regime and the existence of a cross-over.

Figure 3 strongly suggest the existence of three distinct regions. Figure 4 displays the measured domain wall curvature (in log scale) all samples here studied. For samples with thickness ranging between 500μm to 220μm (Fig. 4c) and 180nm to 100nm (Fig. 4a), regions 1 and 3 in Fig. 3, the domain curvature is largely independent of any thickness variations and correlates with the observations in Fig. 3. For samples of thicknesses between 500nm - 200nm, region 2 in Fig. 3, the domain wall curvature increases linearly with sample thickness, Fig. 4b. This corroborates the curvature trends in each thickness region, providing further insight into the monopolar-dipolar regimes and the cross-over region.

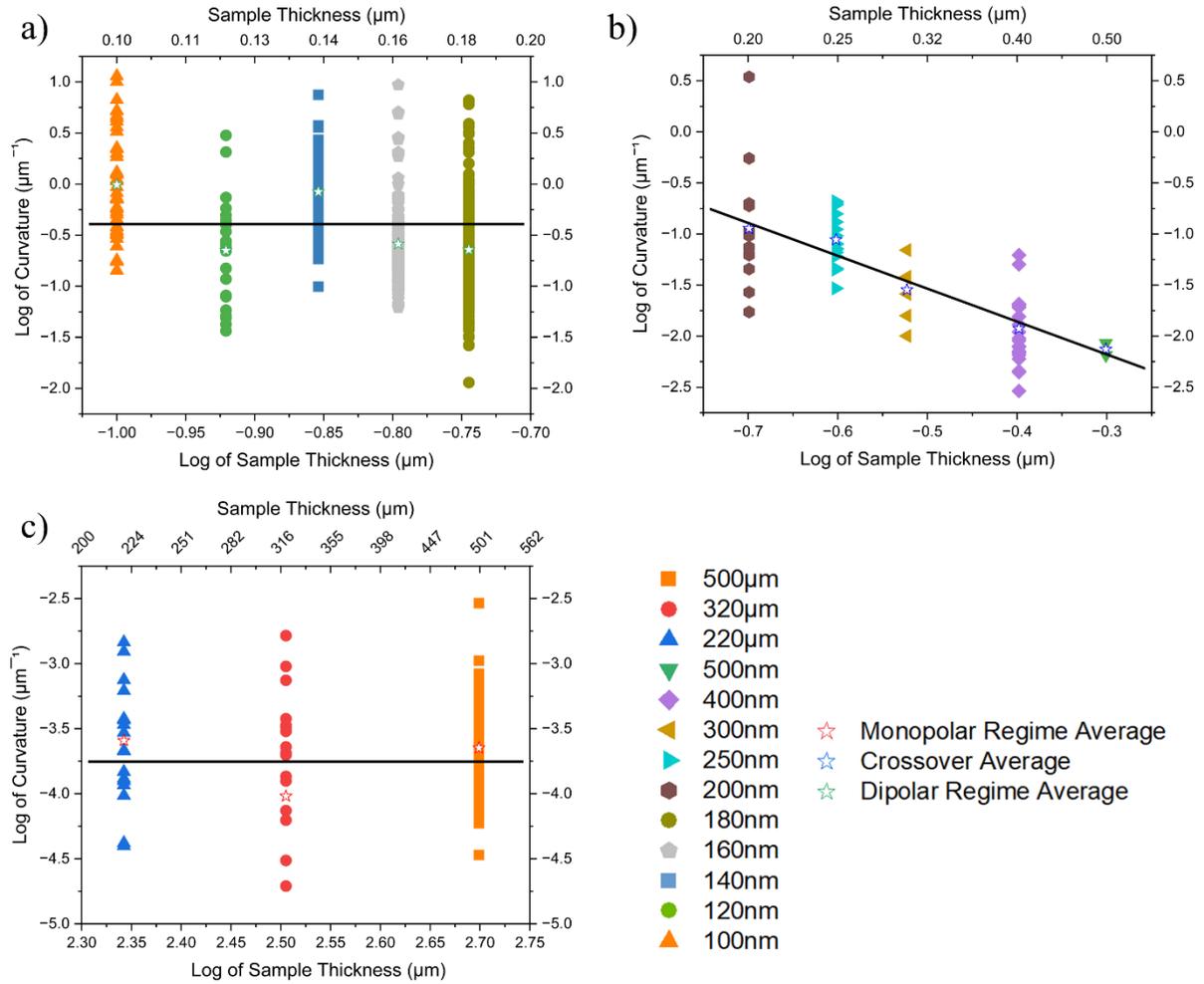

**Figure 4.** Domain wall curvature as a function of sample thickness: **a)** 180nm to 100nm, **b)** 500nm to 200nm and **c)** 500μm to 220μm. Each point corresponds to the measured average curvature per domain wall measured within a given sample thickness. The bulk **a)** and thinnest samples **c)** reveal a flat trend corresponding to the monopolar d$^{-1}$ and dipolar d$^{-2}$ regimes while **b)** shows a close linear trend representing the cross-over.

The increased curvature is a product of the strain generated by kinks within the domain wall, thus it is of interest to measure the local strain around the curved domain wall. This was explored further by scanning precession electron diffraction (SPED).[36] Figure 5a is a virtual bright field image of an area of a 140nm thick sample after heating, here areas with [010]$_{pc}$ straight domain walls and bent [001]$_{pc}$ domain walls are present, marked as i and ii, respectively. The $e_{xy}$ strain and lattice rotation ($\theta$) maps in Figure 5b-c show that the area with straight domain walls (area ii) alternates nicely in sign, resembling previous theoretical work.[26] However, this not the case for the areas with bent domain walls (area ii) where more drastic and localised variations in strain are present.

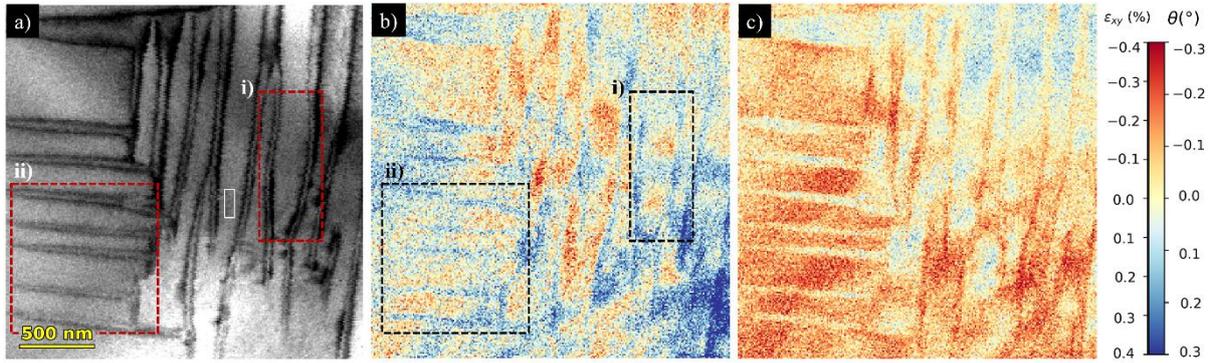

**Figure 5.** SPED strain analysis of a 140nm thick sample after heat cycling. a) virtual bright field image, b) $e_{xy}$ strain along the $[010]_{pc}$ and c) rotation (θ) maps. The white rectangle indicates the reference area and the dotted red squares, marked as area i) and ii), indicate areas of straight and bent domains, respectively. It was noted that the more bent domain walls (region i) display fluctuations of strain, and in turn lattice rotations, along the domain and in the domain walls, respectively. Straighter domain walls (region ii) did not display these changes, except for minor bending of the sample which affects strain and rotation measurements in these areas.

This is further exemplified within a needle domain (Supporting Information Figure 1) where the shear strain dramatically changes towards the needle tip where more pronounced curvature is observed. This would be indicative of the distinct strain levels being induced by the kink interactions, that vary depending on the kink-kink distances, although further experimental work is required to confirm this. A further explanation could be the suppression of the shear strain by the proximity of neighbouring domains.

The physical mechanisms governing this cross-over are further explored by performing atomic simulations (Figure 6) based on a developed (3D) ferroelastic model (see Figure S4). Kinks within a single domain wall are modelled (Figure 6a&c) with focus drawn to kink-kink (Figure 6e) interactions between parallel domain wall (Figure 6b&d). Kink-antikink pairs (Figure 6f) with attracting forces are also considered in our simulations.

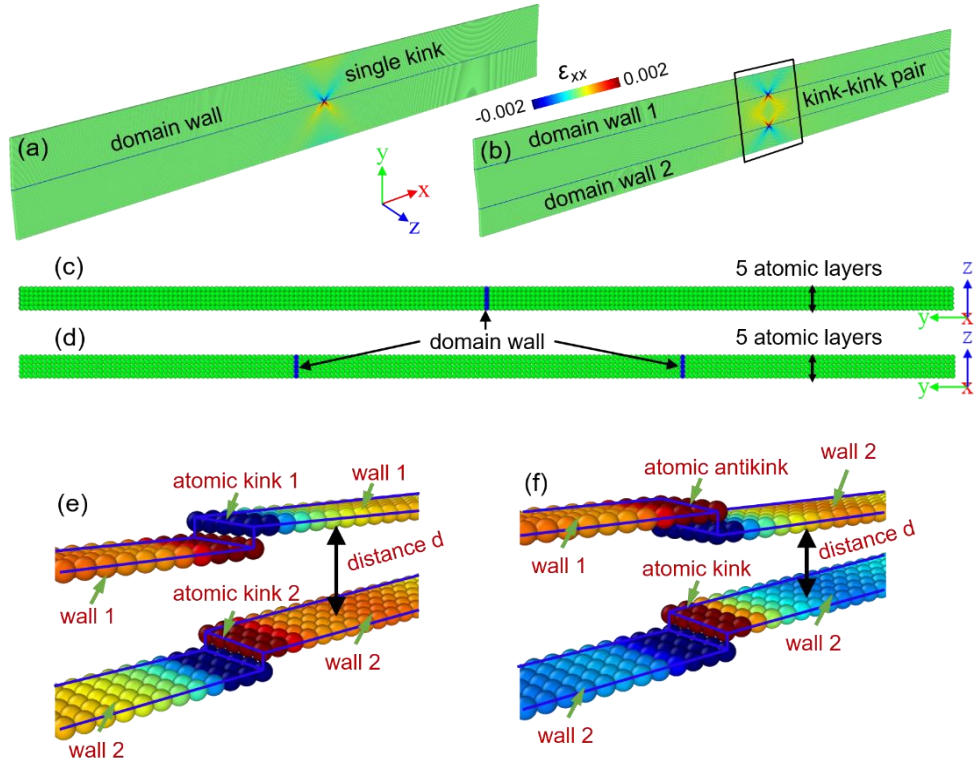

**Figure 6. a)** Static atomic lone kink and **b)** kink pairs containing inside ferroelastic twin domain walls. Interactions between kinks are attractive for kink and anti-kink pairs and repulsive for kink-kink configurations with the latter repelling the configuration away from one other until a reduction in the global energy is met. The simulated 3-dimensional domain wall structure contains 5 atomic layers (c-d) along the z direction while the thickness of the sample varies along the y direction. e) to f) display the strain maps induced by the kink interactions. All strain maps are coded by the atomic-level normal strain, $\varepsilon_{xx}$.

Similar to the experimental approach, we first explore kink induced strain and its effects on the sample thickness. Figure 7 displays the strain and lattice profiles for different samples sizes and wall-wall distances (d) within a 3D model. It shows that above a thickness of 1400 lattice units (l.u.) strain deformations are negligible (Figures 7e, 7-8) however, as the sample size decreases, progressively stronger deformations are observed near the surface (Figures 7e, 1-6). This is exemplified in the lattice profiles (Figures 7a-d) indicating increased bending of the domain walls interacting over larger distances for increasingly smaller sample sizes. The strain profiles simulated here also reflect the observations made in Figure 5, with long ranging strain fields (Figure 5b) emanating from the kinks causing lattice rotations (Figure 5c).

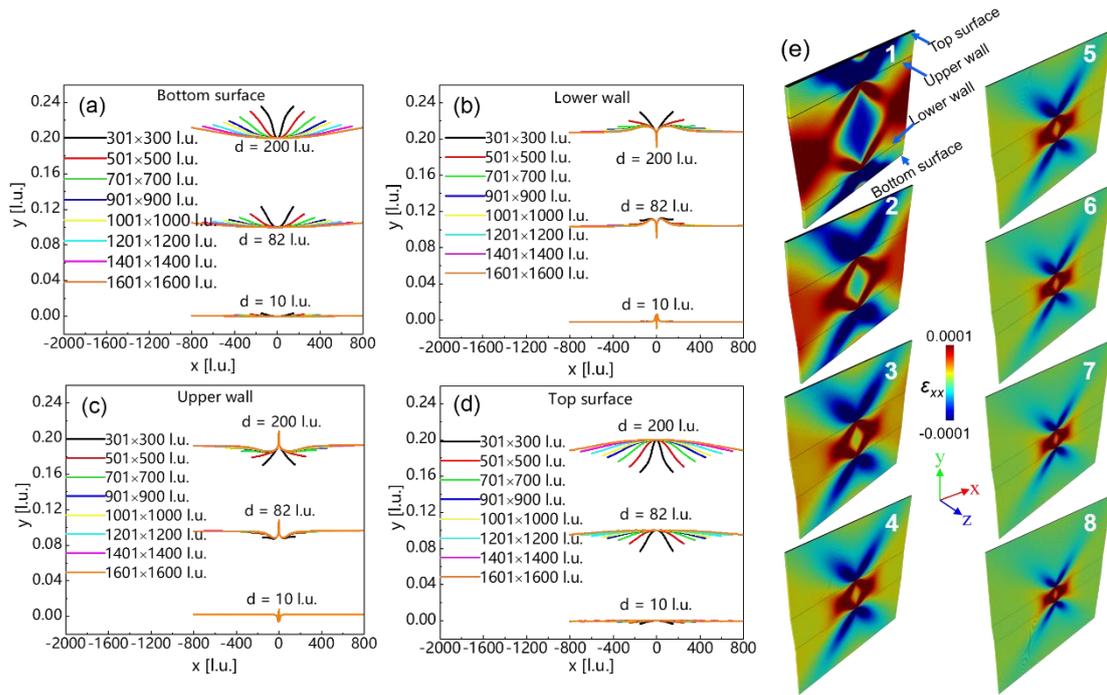

**Figure 7.** Size effects of the macroscopic lattice bending induced by kink-kink interactions with different wall-wall distances *d*. Kink and anti-kink configurations demonstrate the same profiles and relationship. Lattice profiles for the a) bottom surfaces, b) top surfaces, c) lower walls and d) upper walls of samples with *d* = 10 l.u., 82 l.u. and 200 l.u. Strain maps in (e 1-8) are colour-coded according to the atomic-level normal strain: $\varepsilon_{xx}$ with a wall-wall distance of *d* = 200 l.u. The sample sizes in x and y directions in (a-d) are 301 l.u. × 300 l.u., 501 l.u. × 500 l.u., 701 l.u. × 700 l.u., 901 l.u. × 900 l.u., 1,001 l.u. × 1,000 l.u., 1,201 l.u. × 1,200 l.u., 1,401 l.u. × 1,400 l.u., and 1,601 l.u. × 1,600 l.u while all samples are periodic in z direction and contains 5 l.u. The black lines in (e, 1-8) indicate the bottom surface, lower wall, upper wall, and top surface.

The interactions of kink pairs based on sample size is then investigated, Figures 6b&d. A kink positioned centrally in two parallel domain walls is modelled at equivalent distances from the sample midpoint. The interaction energies between kink pairs were simulated in sample sizes ranging from 201x200 l.u. to 1601x160 l.u. with increasing wall-wall distances (Figure 8a). The energy of kink-kink interactions was calculated by reducing the total potential energy by the energy potential of the two noninteracting kinks: $E_{Kink-Kink} + E_{Total} - 2E_{Kink}$. The energy of the kink itself was determined by comparing the difference in potential energy between samples with and without kinks within the domain wall which is dependent on sample thickness.[20] The largest simulated sample sizes demonstrate a monopolar regime with a scaling exponent of -1, Figure 4c. As the sample size is reduced, mirroring the change in sample thickness, the scaling exponents increase until dipolar interactions dominate and further downsizing (from 701x700 l.u. onwards) has no further effect, with a fixed scaling exponent of -2 indicative of the behaviours observed in Figure 4a.

Figure 8b summarises the findings, and the dipolar-crossover-monopolar regimes h as a function of sample size can be clearly identified. Importantly, the change of these exponents is representative of the variation of the surface strain, as seen in Figure 7, which underpins the size-effect dependency observed experimentally.

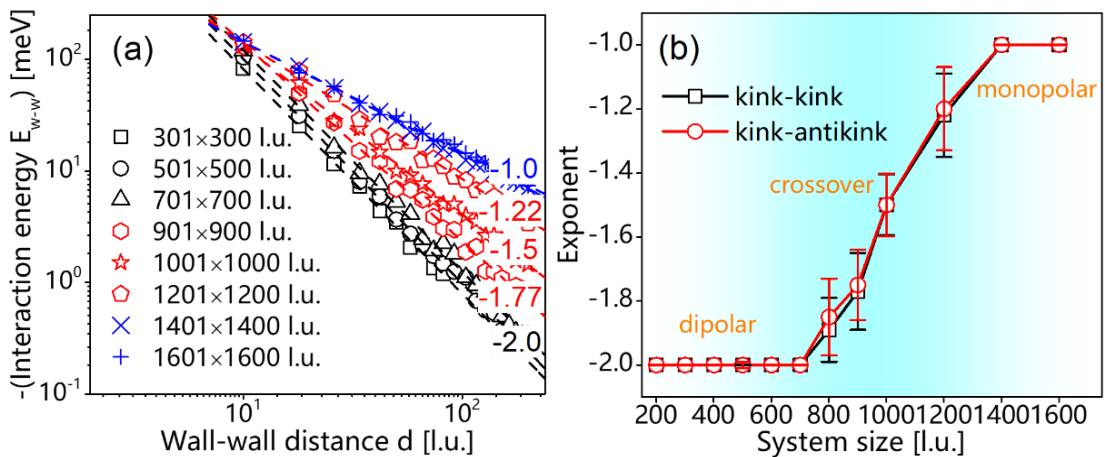

**Figure 8.** Dependence of the scaling law for the kink interactions on the sample sizes. a) Interaction energy on logarithmic scales with the fitted scaling exponents. b) Scaling exponents as a function of sample sizes. The thickness scaling changes from $d^{-2}$ for thin samples to $d^{-1}$ for thick samples.

**Summary of results and discussion**

Experimentally, three regimes are identified in Figure 9. The large samples (Figure 1), resembling bulk samples (thicknesses 500µm to 220µm), exhibit mainly straight domain walls which is indicative of a monopolar regime, governed by junction interactions and domain walls operating over a distance of $d^{-1}$ (Figure 8a). In these thicker samples, the system is unable to relax by bending (Figure 7) and behave in the same manner as samples clamped to a rigid substrate voiding open boundary conditions (Supporting Figure 2) thus reinforcing that bulk like samples hold this relationship (Figure 4c).

As the sample size decreases, a crossover between regimes is distinctly characterised by a linear increase of domain wall curvature between thicknesses of 500nm and 200nm (Figure 4b), along with clustering of domain wall and junction densities (Figure 3, region 2). The experimental results agree reasonably well with interpretations in literature and simulations,[20, 26] alluding to a crossover regime that extends from 350nm to 700nm for a given lattice unit of 0.5nm (Figure 8b).[20, 26] LaAlO$_3$ has a pseudo-cubic lattice parameter with a lower limit of 0.37-0.38nm[37, 38] with some reporting 0.53-0.54nm[39, 40] of which would correspond to an approximate range of 250nm to 600nm and 375nm to 750nm respectively. The lower limit in curvature reported here was found at 200nm (Figures 4a-b), correlating with studies by Lu et. al.[26] that similarly suggest a switch from the crossover to the dipolar regime around 200nm in thickness. If the trends presented in Figure 9 are an indication to the nature of the crossover regime, a suggested critical point from the monopolar regime would be expected at ~1.5µm, which is a magnitude higher than reported in literature[20] and would require further testing.

The smaller the system is, the larger the size effect presented as surface strains on the kink-kink interactions energies (Figures 5&7) which is negligible within bulk like systems. At this point, kinks become the dominant interaction operating over a distance of $d^{-1}$ (Figures 4a&8). Interestingly, the domain and junction density follow a flat trend (Figure 3) suggesting that the two may be intrinsically linked. This is found to be around 180nm within this study (Figure 4a) and theoretically should persist below the 100nm tested here (Figure 8), representative of the dipolar regime.

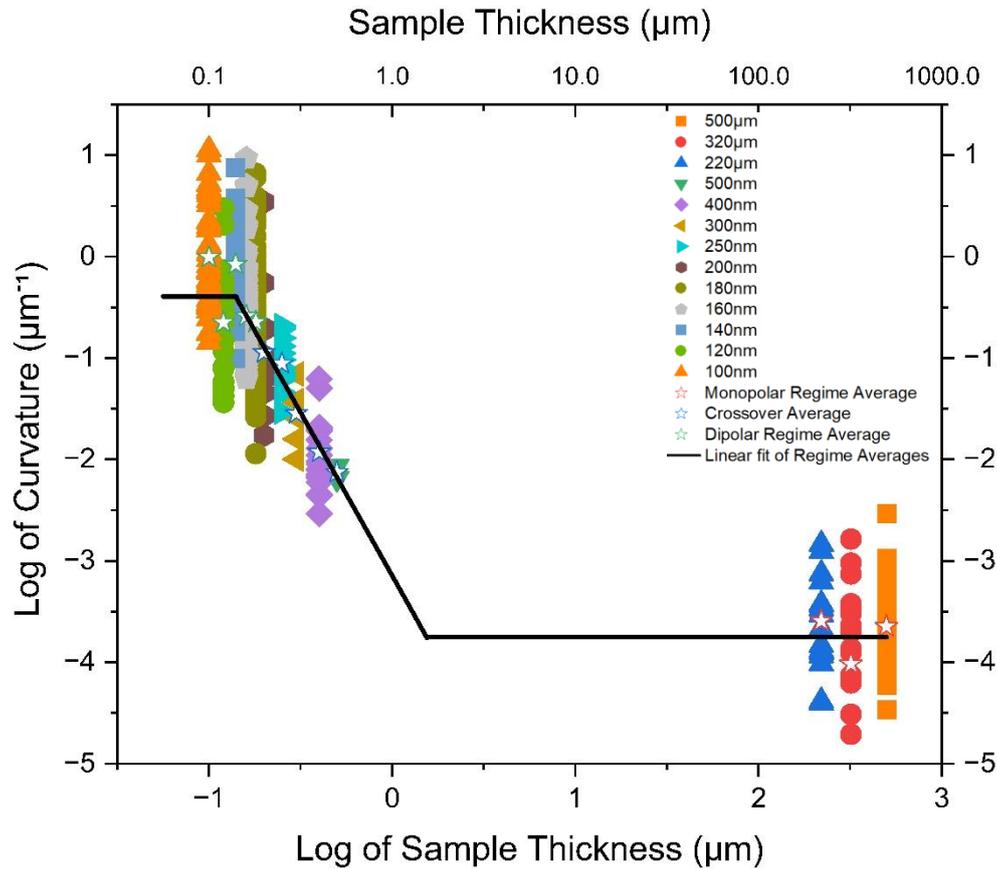

**Figure 9.** Identification of the monopolar, crossover and dipolar regimes. The linear trends for each regime are plotted (solid black line) based on the weighted average of all domain wall curvatures measured for each sample set (thickness). Based off these trends, it is suggested that the monopolar regime exists approximately till 1.5μm where the crossover begins, and the curvature increases linearly. By continuing to decrease the thickness, the existence of the dipolar regime is made clear with a suggested crossover point around 180nm, which persist down to a few unit cells (Figure 8).

**Conclusions**

We present a multiscale study combining in-situ heating microscopy techniques, strain analysis and simulations to study the size effect behaviour of ferroelastic domains. We provide evidence on the behaviour of the monopolar-dipolar regime and their associated crossover.

Samples tested ranged from the bulk (micron scale), which echoes dimensions typically used in wafers etc, down to the nanoscale, reflecting the drive for device miniaturisation as heralded by Moore's law. Specifically, we draw the focus to the mobile atomic steps that exist within domain walls, known as kinks, and their interactions. Simulations of these interactions have been shown to scale differently over a given distance (d) between kinks depending on the sample thickness and manifest as 3 distinct behavioural regimes.

Within the bulk, kink-kink interactions act akin to elastic monopoles in which the interaction energy decays with a $d^{-1}$ dependence from one another. Characteristic of this monopolar regime, domain walls are rigid, and junctions mediate as the predominant interaction mechanism. The densities of these junctions and the domain walls appear to be independent of thickness.

Anticipated to occur around 1-2μm, a crossover regime is present. This regime is broad and has been identified to occur down to approximately 200nm. We characterise this regime based on domain wall curvature which increases linearly with decreasing sample thickness. Physically, in small size samples, surface strain can be expected to affect the kink-kink interactions more significantly, where the kinks inside the domain wall can locally bend the lattice planes and curve lattices with scanning precession electron diffraction (SPED) measurements and complementary simulations providing valuable insight. In more extreme cases within the thinnest samples, it can be expected that domain walls would locally deform the sample, with additional kinks generated within the twin wall. For not so extreme cases, bending in the domain walls might be induced. The bend is more pronounced around the kink with the direction dependent on kink and anti-kink pairing and therefore higher kink concentrations generate greater tilts. The thickness dependence on the type of kink and anti-kink pairing is the same. Although such effects are surprisingly present in thicker samples, the strain involved are not large enough to bend the sample and so remain rigid, making it a good marker of the changes induced by systematically reducing sample thickness.

Finally, reducing the sample thickness to around 180nm see kinks interact via dipolar interactions within the domain walls and the scaling of these exchanges has a $d^{-2}$ dependency. Domain wall and junction densities, as well as domain wall curvature, become thickness independent much like the monopolar regime and mirror the trend of the kink-kink interactions.

Further exploration of the crossover regime and kink dynamics is required where scanning probe and high-resolution TEM techniques should provide valuable insights to the behaviours of this regime in a dynamic manner.

**Methodology**

Lanthanum aluminate (LaAlO$_3$) was selected for this study. As well as being pure-ferroelastic[41] and well documented in literature, this material made an ideal candidate, allowing to bypass the coupling of other ferroic and consider solely the strain effects which commonly mediate the properties of more exotic ferroics and multiferroics.

The samples were prepared in two ways to span the thickness range and test the extent of the monopolar and dipolar regimes. All samples were prepared from the same bulk sample (MTI corp.) to minimise chemical disparities. Bulk samples were cut to size with a diamond wire cutter optimised to apply as little pressure to the sample as possible. Then manually thinned by conventional tripod polishing techniques using a series of polishing papers. These samples were measured optically using an in-plane transmission mode set up and heated using a transmission ready Linkam heating stage, as described elsewhere.[2]

22 free-standing samples with thicknesses ranging from 100nm to 500nm were prepared using a TESCAN Lyra 3 dual beam FIB/SEM, fabricated to have posts either side of the sample, preventing direct contact of the imaging area with the substrate to resemble 'free standing' conditions. The samples were placed on an in-situ heating MEMS chips via conventional ex-situ lift out.[42] Selected thin samples were further thinned by AFM with the technique described further in the Supporting Information S3, before being subjected to in-situ TEM heating. Scanning transmission electron microscopy (STEM) images were recorded simultaneously on high-angle annular dark-field STEM imaging (HAADF) and dark field STEM (DF-STEM) detectors on an FEI TALOS F200 G2 at 200 kV, with a dwell time of 20μs.

All samples were heat cycled in the same manner from room temperature to 600°C (above $T_C$ ~545°C) and back at a rate of 0.33 °C/sec, allowing the domain structure to be reset and select a more optimal configuration. This was selected to mirror the bulk set up, and reducing quenched-in point defects that affect the overall domain patterning.

Domain curvature was measured using the open-source Kappa analysis package [35, 43] in FIJI.[44] A trace of each domain wall was made manually, and a series of curvature points are quantified along the line for each measurable domain wall within the sample. Domain junctions and domain density were measured manually.

All simulations were performed using the Large-scale Atomic/Molecular Massively Parallel Simulator (LAMMPS) code. Ferroelastic domains, domain walls and atomic kinks contained inside walls are described by 3D interatomic potentials based on a landau-type double-well potentials. The interaction potentials have different forms within different atomic inter-distances. Our developed atomic potential is different from the traditional force fields than that of local fine details, such as the atomic kinks are averaged out in traditional potentials. In our model, the atomic kinks contained inside domain walls can be well produced and described, which constitutes these basic fine structures. The model details can be found in the Supporting Information (Figure S4).

Two different boundary conditions are used in our simulations. The first one is the open or Dirichlet boundary condition applied in both x and y directions with periodic boundary conditions in the z direction, where sample relaxations, including shape changes and rotations, are allowed. The second one is more complicated to mimic the bulk effect, i.e., fixed boundary condition of the lower surface and open boundary condition of the upper surface along the y direction, open boundary condition along the x direction and periodic boundary condition along the z direction. Before calculating the kink interactions and lattice bending, all initial structures have been implemented by a long enough energy minimum run to get the fully relaxed structures under each boundary conditions.

**Supporting Information**

Supporting Information is available from the Wiley Online Library or from the author.

## Acknowledgements

JJRS and MA thank Prof. Donald A. MacLaren for facilitating access for the SPED measurements. JJRS and MA is grateful for the support from the EPSRC EP/N509541 grant. GL is grateful for the financial support by the National Natural Science Foundation of China (Grant No. 12304130) and the Doctoral Starting Fund of Yantai University (Grant No. 1115-2222006). EKHS is grateful to EPSRC for support (grant EP/P024904/1). BJR is grateful for the financial support of Foundation Ireland (SFI) and the Sustainable Energy Authority of Ireland (SEAI) under the SFI Career Development Award Grant Number SFI/17/CDA/4637.

## Conflict of Interest

The authors declare no conflict of interest.

## Data Availability Statement

The data that support the findings of this study are available from the corresponding author upon reasonable request.

## Key Words

Domain structure, Ferroelastics, Domain Wall Curvature, surface relaxation, free-standing samples.

# Supporting Information

# Evidence of the Monopolar-Dipolar Crossover Regime: A Multiscale Study of Ferroelastic Domains by In-Situ Microscopy Techniques

John J R Scott[1*], Guangming Lu[2**], Brian J. Rodriguez[3], Ian MacLaren[4], Ekhard K.H. Salje[5], Miryam Arredondo[1]

[1*]School of Mathematics and Physics, Queen's University Belfast, Belfast BT7 1NN, Northern Ireland, UK

[2] School of Environmental and Material Engineering, Yantai University, Yantai 264005, China

[3] School of Physics, University College Dublin, D04 V1W8, Dublin, Ireland

[4] School of Physics and Astronomy, University of Glasgow, Glasgow G12 8QQ, UK

[5] Department of Earth Sciences, University of Cambridge, Cambridge CB2 3EQ, UK

* jscott63@qub.ac.uk ORCID ID: https://orcid.org/0000-0002-7167-2424

** luguangming1990@stu.xjtu.edu.cn

## Supporting Information 1

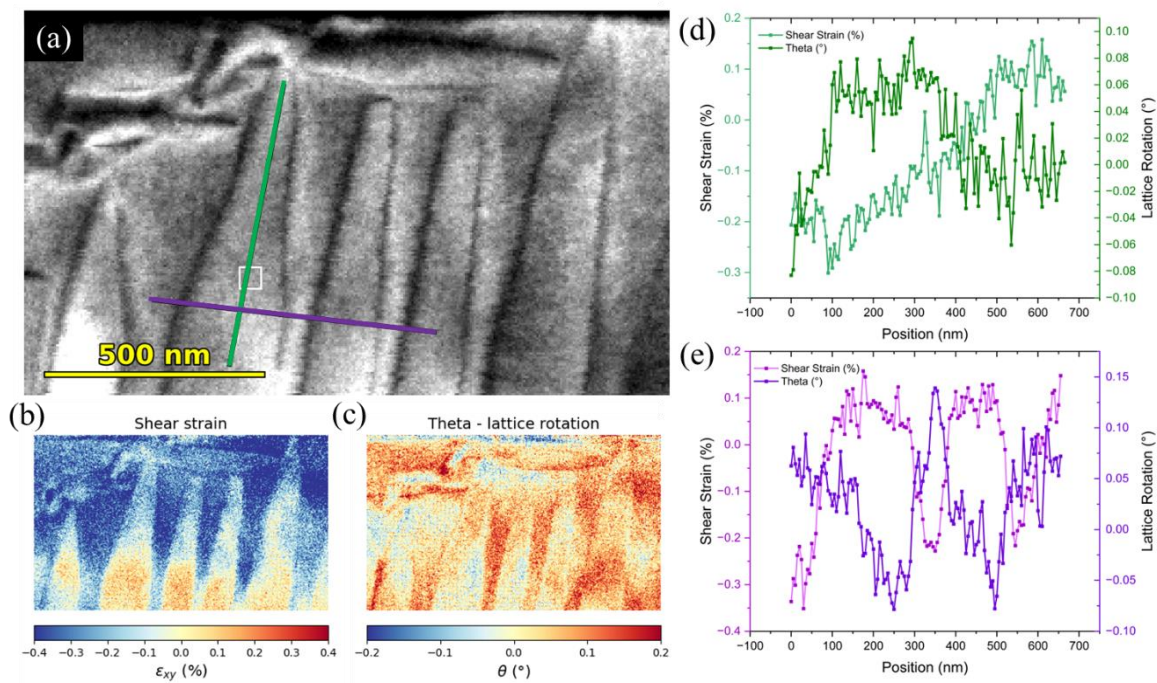

**Figure S1.** Needle domains in a 140nm thick sample. An overview of the needle tips is shown (a) with a strain map (b) and lattice rotation map (c) of the area taken. Line profiles along the needle domain (green) and across several needle domains (purple) are taken in (a), showing the strain and rotation in (d) and (e).

**Supporting Information 2**

Clamped samples, even on one side to a substrate leaving a surface free on the other, thus preventing macroscopic bending (Supporting Figure 2a), demonstrate a $d^{-1}$ monopolar behaviour regardless of sample size. In particular, simulations systematically miss such warping which would make the results of such more realistic for the large majority of ferroelastic materials as well as in the case of porous materials where the similar disorder is observed. These local strain fields produced by kinks, as well as junctions, develop from the discontinuous ferroelastic order parameter across the domain wall which decreases with temperature as described by Landau theory. Supporting Figure 2b shows the lattice displacements associated with a clamped substrate, where only the top, unclamped surface deforms which impacts the interaction between walls (Supporting Figure 2c) (and thus the kinks inside them) which demonstrates a monopolar behaviour which would be akin to the behaviours in a bulk sample (Supporting Figure 2d). This highlights the experimental importance a free-standing sample.

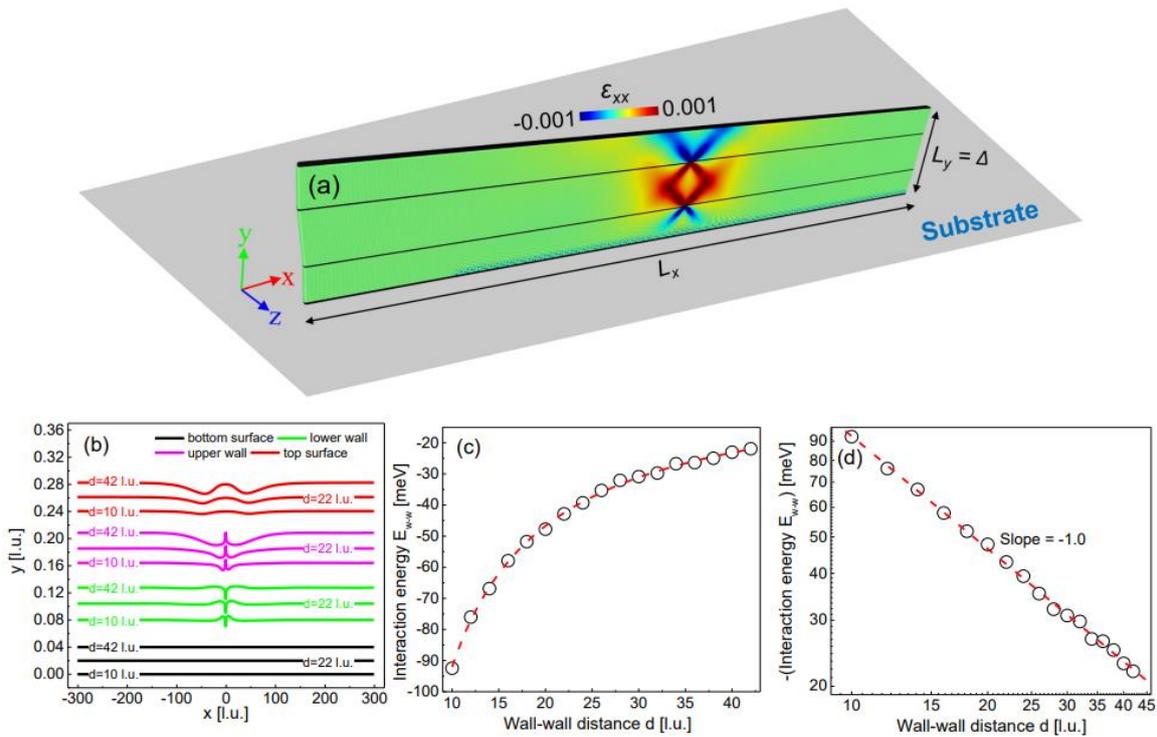

**Figure S2.** Interaction energies of kink-kink configurations with a clamped bottom surface. (a) Strain fields of the thin film with a thickness of $\Delta = 100$ l.u. and wall-wall distances of d = 42 l.u. The strain map was colour-coded according to the atomic-level strain $\varepsilon_{xx}$. (b) Lattice displacements of fixed bottom surface, lower wall, upper wall, and top surface due to the kink-kink interactions. (c) The variation of interaction energies as a function of wall-wall distance. The data points in (c) are fitted by using the equation $E_{kink-kink} = E0 - A \times d^B$ with $E0 \sim 0$ (noninteracting kinks), $A = 0.832$ eV l.u. and $B = -1$. The scaling exponent of -1 is shown in (d).

**Supporting Information 3**

Selected free-standing samples were further thinning by using an AFM. Firstly, the sample was clamped with ion beam deposited platinum welds on the posts either side of the sample. The sample was first located using amplitude modulation-mode AFM and then the surface was 'etched' using contact-mode AFM, in the direction indicated by the green arrow in Figure S3. The sample surface was then cleaned using contact-mode AFM with a lower applied force to scan debris out of the field of view and the heat cycle was subsequently applied, as described in the methodology. Importantly, it was noted that the AFM thinning did not alter the domain pattern significantly.

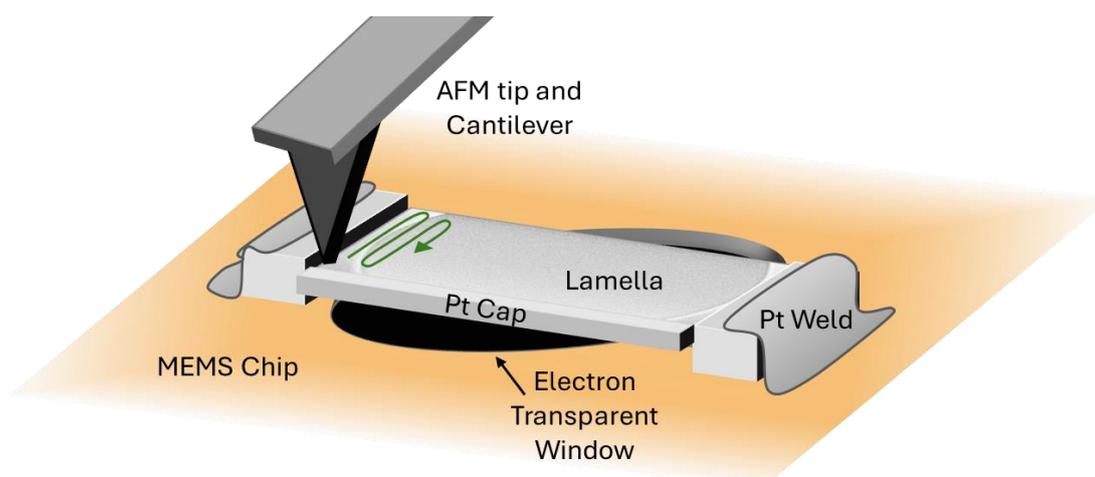

**Figure S3.** Schematic of lamella thinning via AFM milling.

**Supporting Information 4**

The ferroelasticity (anharmonic landau interactions) is confined inside 2-dimensional xy plane (Figure S4(b)) with three different potential energy terms: the harmonic first nearest atomic interactions $U(r) = 20(r - 1)^2$ (black springs), the anharmonic second-nearest interactions $U(r) = -25(r - \sqrt{2})^2 + 20000(r - \sqrt{2})^4$ (red springs) along diagonals in the lattice unit and the fourth-order third-nearest interactions $U(r) = 8(r - 2)^4$ (red arrows), where r is the distance between atoms. The constructions of the landau springs (red springs in Figure S4(b)) are inspired by a well-known ferroelastic phase transition of SrTiO$_3$ with the energy scale determined by T$_c$= 105K and a typical ferroelastic shear angle of 2°. Different atomic layers in the z direction, are connected by two harmonic springs (springs in Figure S4(c)) with potential energy forms of $U(r) = 20(r - 1)^2$ for first-nearest interactions (orange springs) and $U(r) = 2(r - \sqrt{2})^2$ for second-nearest interactions (green springs). The harmonic springs constitute the elastic background for the ferroelastic model while fourth-order and third-nearest interactions in Figure S4(b) are constructed to keep reasonable domain wall thicknesses. The atomic mass is M = 50 amu.

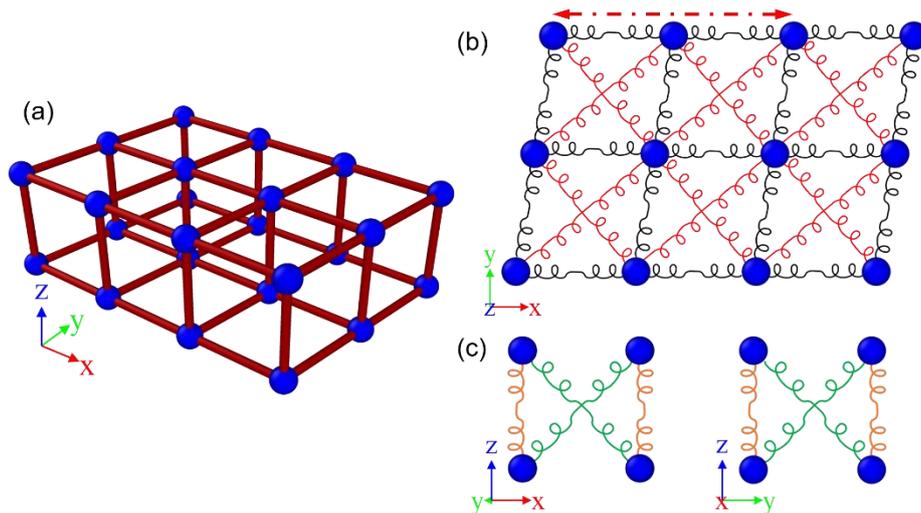

**Figure S4**. 3-dimensional interatomic potential for a generic ferroelastic model (a). Ferroelasticity is confined to the xy plane (b) consisting of three interatomic interactions, i.e., nearest-neighbour (black springs), next-nearest-neighbour (red springs), third-nearest-neighbour (red arrow). The red springs are Landau springs with a double well potential so that the energy has a minimum value when the lattice is sheared with respect to the cubic unit cell. The atomic layers in xy plane are connected in z direction by two harmonic springs (orange and green springs in (c)).